\documentclass[twocolumn,showpacs,preprintnumbers,amsmath,amssymb,floatfix]{revtex4}

\usepackage{epsfig,psfrag}
\usepackage{dcolumn}
\usepackage{bm}
\usepackage{graphicx}
\usepackage{pstricks}
\usepackage{color}
\usepackage{verbatim}

\begin{document}
\title{Iterative real-time path integral approach to  
nonequilibrium quantum transport}
\author{S. Weiss, J. Eckel, M. Thorwart and R. Egger}
\affiliation{Institut f\"ur Theoretische Physik,
Heinrich-Heine-Universit\"at D\"usseldorf, 40225 D\"usseldorf, Germany}
\date{\today}
\begin{abstract}
We have developed a numerical approach to 
compute real-time path integral expressions for 
quantum transport problems out of equilibrium.  
The scheme is based on a deterministic iterative summation of 
the path integral (ISPI) for the generating
function of the nonequilibrium current. 
Self-energies due to the leads, being non-local in time, 
are fully taken into account within a finite memory time, 
thereby including non-Markovian effects, and numerical results are
extrapolated both to vanishing (Trotter) time discretization
and to infinite memory time.  This extrapolation scheme converges except at 
very low temperatures, and the results are then numerically exact.  
The method is applied to 
nonequilibrium transport through an Anderson dot.
\end{abstract}
\maketitle

\section{Introduction}
\label{intro}

Quantum transport has attracted theoretical and 
experimental research since it offers the possibility to 
investigate quantum many-body properties at as well as out of 
thermodynamic equilibrium 
\cite{Heinzel}. 
The ongoing improvement in miniaturization down to the nanometer scale 
allows to study electron transport in ultra-small devices, e.g., 
in single molecules or artificially designed quantum dots 
\cite{molel,Nitzan07,Weber02,Ruitenbeek97,Scheer04,Park}. 
There is a broad variety of interesting physical effects, due to 
interactions or the nonequilibrium conditions arising when a bias 
voltage is applied to the source and drain electrodes \cite{Koenig,Meir,Hershfield}. These range from Coulomb 
blockade via coherent  (e.g., resonant tunneling) transport
to the Kondo effect, to name but a few. While on the 
experimental side, progress stems from an increased control 
of fabrication processes, many theoretical works deal with 
refined approximation schemes applicable in different parameter 
regimes. However, {\em exact\/} theoretical results --- either analytical or 
numerical --- for nonequilibrium quantum 
transport systems are rare, mainly because of the lack of adequate methods allowing to tackle such questions. There clearly is a 
considerable need for numerically exact methods to describe 
nonequilibrium quantum transport, both to check analytical (and usually approximate) approaches and to connect theory to experiment. 
Here, we mean by
nonequilibrium transport specifically those phenomena which go beyond the
standard approach of linear response to the applied bias voltage. 

In this work, we propose a novel numerical scheme denoted as 
{\sl iterative summation of real-time path integrals}\ (ISPI), in order to 
address quantum transport problems out of equilibrium.
Many-body systems driven out of equilibrium are known
\cite{Abanin05,Kaminski00,Mitra05b} to acquire a steady state that may
be quite different in character from their ground-state properties.
Details of the steady state may depend on the nature of the
correlations, as well as on the way in which the system is driven out
of equilibrium. While there are a variety of nonperturbative
techniques in place to study equilibrium systems, many of these
methods cannot be extended to nonequilibrium systems in a
straightforward way. 

Different approximations have been pursued 
previously in order to tackle nonequilibrium situations. 
For instance, transport through an Anderson dot in the
Kondo regime has been described theoretically along several different lines,
e.g., for the asymptotic low-energy regime by Fermi liquid theory \cite{Oguri01},
via interpolative
schemes \cite{Aligia},  using integrability concepts \cite{Konik02}, 
or by the perturbative renormalization group \cite{Kaminski00,Rosch04}.
Transport features of the Anderson model
have also been discussed by perturbation theory in the interaction strength
\cite{Fujii,rubio}.  Sophisticated nonperturbative methods have been
developed in order to extract exact results out of equilibrium 
for special models, where integrability is available. 
For instance,  the interacting resonant level model  presently
enjoys much interest 
\cite{Komnik,Doyon1,Mehta,Borda,Doyon2,Boulat,Nishino}. 
Universal aspects of nonequilibrium currents in a quantum dot
have been discussed in Ref.\ \cite{Doyon1}. In various perturbative
regimes, this model has been addressed by field theory
methods \cite{Borda,Doyon2,Boulat,Nishino}. 
However, the underlying theoretical concepts are still under much debate,
and partially conflicting results were reported in the literature,
cf.~Refs.  \cite{Borda,Doyon2,Boulat}.

The above discussion shows that precise numerical simulation tools
are in great demand in this field.  Let us briefly review the existing
numerical methods available for nonequilibrium transport.
Much effort has been directed to the application of renormalization group 
(RG) techniques to this problem.  For instance, a perturbative real-time RG 
analysis of nonequilibrium transport has been performed \cite{Schoeller}, 
and steps towards the appropriate generalization of the
 density matrix renormalization group technique have appeared  
\cite{Schmitteckert,Schollwoeck}.  In addition, the functional 
RG approach has been generalized to non\-equi\-li\-bri\-um 
recently \cite{Jakobs}. It forms a systematic and fast 
perturbative expansion scheme, yielding reliable results 
for small interaction strengths, where the infinite hierarchy of
 equations for the self-energy can be cut after the first few steps. Moreover, a
possible extension of Wilson's numerical RG approach to 
nonequilibrium has been discussed \cite{Costi1,anders}.
Another line of attack considered 
numerically exact  real-time quantum Monte Carlo (QMC) simulations 
 (for the corresponding equilibrium case, see Refs.\ 
\cite{Oguri95,Wang}). Due to the sign problem, however, these calculations 
become increasingly difficult at low temperatures. 
Refined multi-level blocking techniques \cite{Egger00,Egger03} 
allow to achieve further progress, but numerical simulations remain
hard within this approach. 
(For recent progress, however, see Ref~\cite{lothar}.) 
Let us also mention the flow equation method, which has been 
applied to study the nonequilibrium Kondo effect \cite{kehrein}. 
Finally, a very recent development considers non-standard
ensembles to describe steady state transport, but involves
a numerically troublesome analytic continuation \cite{Han}.

Our ISPI approach, described in detail below, provides a novel alternative, 
and in principle numerically exact, method to tackle out-of-equilibrium 
transport through correlated quantum dots. 
The method is deterministic and, hence, there is no sign problem. 
It is based on the evaluation of the full 
nonequilibrium Keldysh generating function, including suitable source
terms to generate the observables of interest. 
 It builds on the fact that the fermionic leads induce self-energies 
that are non-local in time, but which decay 
exponentially in the long-time limit at any 
finite temperature. 
Thus, a memory time exists 
such that within this time span, the correlations are exactly taken 
into account, while 
for larger times, the correlations 
can be dropped. This allows to construct an iterative scheme to
evaluate the generating  function. An appropriate 
extrapolation procedure allows then to eliminate both the 
Trotter time discretization error 
(the Hubbard-Stratonovich (HS) transformation 
below requires to discretize time), and 
the finite memory-time error, yielding  
finally the desired numerically exact value for
the observables of interest. 
 The extrapolation procedure is convergent for not too low 
temperatures, since then memory effects are exponentially small for long times.
If the extrapolation converges, we thereby obtain numerically exact results 
for nonequilibrium quantum transport properties of interacting systems. 

The ISPI scheme is implemented here for the example of the 
 well-known single-level Anderson impurity model 
\cite{Anderson,Flensberg,Tsvelik,Schiller,Horvath},
 but appropriately modified, it can be applied to other quantum dot models as well.
The nonequilibrium current is calculated from a generating function,
represented as a real-time path integral in the Keldysh formalism. 
After a Hubbard-Stratonovich transformation 
employing an auxiliary Ising spin field, all fermionic degrees of
freedom (of the dot and the leads) can be integrated out
exactly, however, at the price of introducing time-nonlocal self-energies 
 for the leads and a path summation over all Ising spin configurations. 
For this, an iterative summation scheme is constructed, exploiting  that 
the time-nonlocal correlations in the lead self-energy effectively decay exponentially at finite temperature $T$ or bias voltage $V$, thereby setting the 
characteristic memory time.  
For larger times, the correlations are exponentially
small and will be neglected. The full time-discretized Keldysh Green's function 
(GF) of the dot then assumes a band matrix structure, where the
determinant can be calculated iteratively using its Schur form. The
scheme is constructed such that within the range set by the memory 
time, the  path integral is evaluated exactly. The remaining
systematic errors are the Trotter time discretization and the 
finite-memory error. Both, however, can be eliminated in a 
systematic way based on a Hirsch-Fye type extrapolation procedure. 
Based on the above construction principle, our approach cannot be
applied at very low energies ($T,V\to 0$), where, fortunately, other
methods are available.
For finite temperatures, the requirement of not too long memory times 
can be met, and the spin path summation remains tractable. 

The present paper is organized as follows. In Sec.\ \ref{sec:Model}, we 
introduce the model for the quantum dot coupled to normal 
leads. We present the computation of the generating  
function for the nonequilibrium Anderson model in Sec.\ \ref{sec:Keldysh} 
in the presence of an external source term, which allows to 
calculate the current as functional derivative. In Sec.\ \ref{sec:IPI} 
we introduce the numerical iterative path-integral summation 
method, from which we obtain observables of interest. We give a 
detailed discussion of the convergence properties of our method 
 and describe the extrapolation 
scheme in Sec.\ \ref{sec:extrapol}. 
For several sets of parameters, we will present  
results in Sec.\ \ref{sec:Results}, followed 
by a discussion  in Sec.\ 
\ref{sec:Conclusion}. Throughout the paper, we set $\hbar=k_B=1$. 
 
\section{Model}
\label{sec:Model}

\begin{figure}[t!]
\vspace{0.2cm}
\includegraphics[width=80mm]{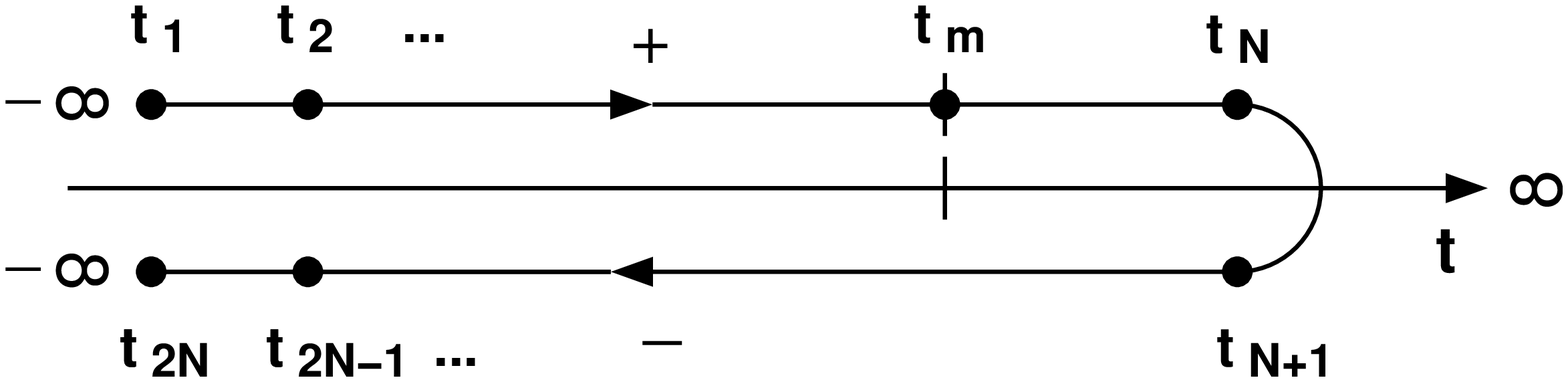}
\caption{ \label{Keldysh} Keldysh contour: every physical 
time has two contour representatives on the branches 
$\pm$. The measurement time $t_m$ only has a single representative 
on the upper branch.}
\end{figure}

We consider the Anderson model \cite{Anderson} given by the Hamiltonian 
\begin{eqnarray}
\mathcal{H}&=&H_{dot}+H_{leads}+H_T \nonumber\\
&=&\sum_\sigma E_{0\sigma} \hat n_\sigma
+U \hat n_\uparrow \hat n_\downarrow + 
\sum_{kp\sigma}(\epsilon_{kp}-\mu_p)c^\dag_{kp\sigma}c_{kp\sigma}\nonumber\\
&&- \sum_{kp\sigma} \left[t_p c_{kp\sigma}^\dagger d_\sigma + h.c.\right].
\label{andham}
\end{eqnarray}
Here, $E_{0\sigma}=E_0+ \sigma B$ with $\sigma=\uparrow,
\downarrow = \pm$ is the  energy of a single electron with
spin $\sigma$ on the isolated 
dot, which can be varied by tuning a back gate
voltage or a Zeeman magnetic field term $\propto B$. The latter is 
assumed not to affect the electron dispersion in the leads. 
The corresponding dot electron annihilation/creation operator is 
$d_\sigma/d_\sigma^\dagger$, with  $\hat n_{\sigma}\equiv d^\dag_\sigma
d_\sigma$ with eigenvalues $n_\sigma=0, 1$, and  
 $U$ denotes the on-dot   interaction.  
For later purpose, it is  convenient to use the operator identity 
$\hat n_\uparrow \hat n_\downarrow = \frac{1}{2}(\hat n_\uparrow + \hat n_\downarrow)
-\frac{1}{2}(\hat n_\uparrow - \hat n_\downarrow)^2 $,
thereby introducing the shifted single-particle 
energies $\epsilon_{0\sigma}\equiv E_{0\sigma}+U/2,$
which yields the equivalent dot Hamiltonian 
\begin{equation}
H_{dot}=H_{dot,0}+H_U=
\sum_\sigma \epsilon_{0\sigma} \hat n_\sigma 
- \frac{U}{2}(\hat n_\uparrow-\hat n_\downarrow)^2  . 
\end{equation}
In  Eq.\ (\ref{andham}), $\epsilon_{kp}$ denotes 
the energies of the noninteracting 
electrons (operators $c_{kp\sigma}$) 
in lead $p=L/R=\pm$, with chemical potential $\mu_p=p eV/2$. 
Dot and leads are connected by the tunnel couplings $t_p$. 
The observable of interest is the (symmetrized) tunneling 
current $I=(I_L-I_R)/2$,
\begin{equation}
I(t)=-\frac{ie}{2}\sum_{kp\sigma}\left[pt_p \langle c_{kp\sigma}^\dag d_\sigma\rangle_t
-pt_p^* \langle d^\dag_\sigma c_{kp\sigma}\rangle_t
\right],
\end{equation}
where $I_p(t) = -e \dot{N}_p(t)$ with $N_p(t)= 
\langle \sum_{k\sigma}c_{kp\sigma}^\dagger 
c_{kp\sigma} \rangle_t$. The stationary steady-state dc current follows as the
asymptotic long-time limit, $I=\lim_{t\to\infty}I(t)$. 
We have explicitly confirmed that current conservation, $I_L+I_R=0$, 
is numerically fulfilled  for the ISPI scheme.
In the presence of a finite bias voltage, $V\neq 0$, the Keldysh 
technique \cite{Keldysh,Rammer,Kamenev} provides a way 
to study nonequilibrium transport. In this formalism, 
the time axis is extended to a contour with $\alpha=\pm$ branches, 
see Fig.\ \ref{Keldysh},
along with an effective doubling of fields. The Keldysh GF is 
$G^{\alpha\beta}_{ij}(t_\alpha,t'_\beta)=-i\langle \mathcal{T}_C[\psi_i(t_\alpha)
\psi^\dag_j(t'_\beta)]\rangle,$
where $\mathcal{T}_C$ denotes the contour ordering of times along the 
Keldysh contour, 
and $i,j=L,R,0$ correspond to  fields 
representing lead or dot fermions, respectively. 
We omit the spin indices here, remembering that 
each entry still is a diagonal $2\times 2$ matrix in spin space. 
In general, the four Keldysh components are linearly dependent, such 
that $G^{++}+G^{--}=G^{+-}+G^{-+}$. However, the commonly used Keldysh rotation 
\cite{Kamenev, Rammer} seems to offer no advantages here, and is not 
employed in what follows.

\section{Generating function}
\label{sec:Keldysh}

Let us then discuss the generating function, which contains all
relevant information about the physics of the system. 
 First, we want to integrate 
out the lead fermion fields, and, in addition, perform a
discrete Hubbard-Stratonovich transformation. 
This allows  to integrate out the dot fields as well,  
and we are then left with a discrete
path summation. We start with the fermionic path-integral representation of
the generating function, 
\begin{equation}
Z[\eta]=\int \mathcal{D}\left[\prod_\sigma \bar{d}_\sigma , d_\sigma ,
\bar{c}_{kp\sigma} ,
c_{kp\sigma}\right]e^{iS[\bar{d}_\sigma, d_\sigma ,
\bar{c}_{kp\sigma}, c_{kp\sigma}]} ,
\end{equation} 
with Grassmann fields  $(\bar{d},d,\bar{c},c)$ 
(Note that we use the same symbols for fermion operators and Grassmann fields 
throughout the paper for better
readability). We 
 introduce  an external source term $S_\eta$ (defined 
 below in Eq.\ (\ref{fullact})), which allows to compute the 
current at measurement time $t_m$, 
\begin{equation} \label{gencurrent}
I(t_m)=\left.-i\frac{\partial}{\partial\eta}\ln Z[\eta]\right|_{\eta=0}.
\end{equation}
We note in passing that it is also possible to evaluate other 
observables, e.g., the zero-frequency noise, by introducing appropriate source
terms and performing the corresponding derivatives. 
The action is $S=S_{dot}+S_{leads}+S_T+ S_{\eta}$, where 
\begin{eqnarray} \label{fullact}
S_{dot} & = & S_{dot,0} + S_U \nonumber \\ 
&=&\int_C dt \left[\sum_\sigma \bar{d}_\sigma
\left(i\partial_t-\epsilon_{0\sigma}\right)d_\sigma
+\frac{U}{2}(n_\uparrow -n_\downarrow)^2\right], \nonumber\\
S_{leads}&=&\int_C dt \sum_{kp\sigma}\bar{c}_{kp\sigma}
(i\partial_t-\epsilon_{kp}+\mu_p)c_{kp\sigma}, \nonumber\\
S_{T}&=&\int_C dt \sum_{kp\sigma}t_p
{\bar c}_{kp\sigma}d_\sigma+h.c.,\nonumber\\
S_{\eta}&=&
\frac{ie\eta}{2}\sum_{kp\sigma}p(t_p\bar{c}_{kp\sigma}
d_\sigma-t^*_p\bar{d}_\sigma c_{kp\sigma})(t_m).
\end{eqnarray}
The interaction term $S_U$ in Eq.\ (\ref{fullact}) does 
not allow to directly perform the functional integration.  Apart from this, 
all other terms are quadratic in the Grassmann fields and thus  define 
 the `noninteracting' (quadratic) part. Note that the level shift
 $+U/2$ is here included in the noninteracting sector. 

\subsection{Noninteracting part}
\label{sec:nonin}

Before turning to the interacting problem, we briefly discuss the Keldysh 
GF of the noninteracting problem. 
Let us first integrate out the lead degrees of freedom. We 
remain with an effective action for the dot, which is non-local 
in time due to the presence of the leads. In particular, after 
Gaussian integration, the generating function   reads
\begin{eqnarray}
Z_{ni}[\eta]=\int \mathcal{D}\left[\prod_\sigma \bar{d}_\sigma , d_\sigma
\right]e^{i(S_{dot,0}+S_{env})}
\end{eqnarray}
with
\begin{eqnarray}
S_{env}&=& \int_Cdt\int_C dt' \sum_\sigma\bar{d}_\sigma(t)
\bigg\{\gamma_L(t,t')+\gamma_R(t,t')\nonumber\\
&+&\frac{ie\eta}{2}[\gamma_L(t,t')-\gamma_R(t,t')] \nonumber \\
&\times & 
[\delta(t-t_m)-\delta(t'-t_m)]\bigg\}d_\sigma(t').
\label{Seff} 
\end{eqnarray}
For the source term, the physical measurement time $t_m$ 
can be taken at one branch of the contour, see Fig.\ \ref{Keldysh}. 
As we fix $t_m$ on the upper $(+)$ branch, the $(--)$ Keldysh 
element of the source term self-energy vanishes, see the time-discretized 
version below. 
The $\gamma_p$ in Eq.\ (\ref{Seff}) 
represent the  leads, and their Fourier transforms are explicitly 
given as $2\times 2$ Keldysh matrices, 
\begin{equation}
\gamma_p(\omega)=i\Gamma_p\left(
\begin{array}{cc}
2f(\omega-\mu_p)-1&-2f(\omega-\mu_p)\\
2-2f(\omega-\mu_p)& 2f(\omega-\mu_p)-1
\end{array}\right). 
\end{equation}
Here, we have used the fact that the 
 leads are in thermal equilibrium,  $f(\omega)=1/(e^{\omega/T}+1)$. 
Moreover, we take the standard wide-band limit, with 
a constant density of states per spin channel around the Fermi 
energy, $\rho(\epsilon_F)$, yielding the 
hybridization $\Gamma_p=\pi\rho(\epsilon_F)|t_p|^2$ of the dot level
with lead $p$. For the sake of clarity, we assume from now on
symmetric contacts,
 $\Gamma_L=\Gamma_R\equiv \Gamma/2$.
The generalization to asymmetric contacts is straightforward. 
In  the next step, still for $U=0$, we may also integrate over the 
dot degrees of freedom. 
We obtain the noninteracting generating function 
\begin{equation}
Z_{ni}[\eta]= \prod_\sigma \det \left[
-i G_{0\sigma}^{-1}(t,t')+\eta \Sigma^J(t,t') 
\right] \, ,
\end{equation}
where $G_{0\sigma}^{-1}(t,t')$ follows from 
\begin{eqnarray}\label{fullD}
G_{0\sigma}(\omega) & = & 
\left[(\omega-\epsilon_{0\sigma})\tau_z-\gamma_L(\omega)-\gamma_R(\omega)\right]^{-1}
\nonumber \\
& = & \frac{1}{\Gamma^2+(\omega-\epsilon_{0\sigma})^2}\nonumber\\
&\times&\left( \begin{array}{cc} \omega-\epsilon_{0\sigma}+i\Gamma(F-1)&i\Gamma F\\
i\Gamma(F-2)&-\omega+\epsilon_{0\sigma}+i\Gamma(F-1) \end{array} \right)\, \nonumber
\end{eqnarray}
 where  
$\tau_z$ is the standard Pauli matrix in Keldysh space, 
and $F=f(\omega+eV/2)+f(\omega-eV/2)$.  Moreover, the self-energy 
for the source term is obtained as
\begin{eqnarray}
\Sigma^J (t, t') &=& \frac{e}{2} \left[\gamma_L(t, t') - 
\gamma_R(t, t')\right] \nonumber \\
& \times & \left[ \delta(t-t_m)-\delta(t'-t_m)\right] \, .
\end{eqnarray}
Up to this point, we have discussed the noninteracting case. 
In order to treat interactions, we now 
 perform a Hubbard-Stratonovich transformation.

\subsection{Hubbard-Stratonovich transformation}

Let us therefore turn back to the real-time action $S_{dot}$ of 
the dot in the presence of interactions, $U\ne 0$, see Eq.\ (\ref{fullact}). 
For later numerical purpose, it is beneficial to proceed with 
the discussion in the time-discretized representation of the 
path integral \cite{negele}. In order to decouple the quartic term, we 
discretize the full time interval, $t = N \delta_t$, with the 
time increment $\delta_t$. On each time slice, we perform a Trotter 
breakup of the dot propagator according to 
$e^{i\delta_t(H_{0}+H_T )} = e^{i\delta_t H_T/2}e^{i\delta_t H_{0}}
e^{i\delta_t H_T/2}+O(\delta_t^2)$,
where
$H_0=H_{dot}+H_{leads}$. By this, we introduce a Trotter error 
\cite{HirschFye,Hirsch,Fye} which, however, 
will be eliminated from the results in a systematic way \cite{deRaedt,Eckel}, 
see below.
Next, we use a discrete Hubbard-Stratonovich transformation 
\cite{Hirsch,Hubbard,Siano} for the interacting part, which introduces 
Ising-like discrete spin fields $s_n^\pm =(s^+_n, s^-_n)$ 
on the $\alpha=\pm$ branches of the 
Keldysh contour (with $s^\alpha_n = \pm 1$) on  the $n$-th Trotter slice. 
For a given Trotter slice, we now use 
\begin{equation}
e^{\pm i\delta_tU(\hat n_\uparrow-\hat n_\downarrow)^2/2}=
\frac{1}{2}\sum_{s^\pm=\pm}e^{-\delta_t\lambda_\pm s^\pm(\hat n_\uparrow-\hat n_\downarrow)}.
\end{equation}
For $U>0$, noting that $n_\uparrow-n_\downarrow=0, \pm 1$, the sum can be
carried out and gives the condition
\begin{displaymath}
\cosh(\delta_t\lambda_\pm)=\cos(\delta_tU/2)\pm i\sin(\delta_tU/2).
\end{displaymath}
The solution is $\lambda_\pm=\lambda'\pm i\lambda''$ with 
\begin{equation}
\delta_t\lambda'=\sinh^{-1}\sqrt{\sin(\delta_tU/2)}, 
\, \, \, \delta_t\lambda''=\sin^{-1}\sqrt{\sin{(\delta_tU/2)}}.
\end{equation}
Note that the overall sign of $\lambda_\pm$ 
is chosen arbitrarily, but does not 
influence the physical result. Uniqueness of  this 
Hubbard-Stratonovich transformation requires  $U\delta_t<\pi$.
To ensure sufficiently small time discretizations, in all
calculations one should then obey the condition
${\rm max}(U, e|V|, |\epsilon_0|, T) \lesssim 1/\delta_t$.
 
\subsection{Total GF and generating function}

After the Hubbard-Stratonovich transformation, 
the remaining fermionic Grassmann variables $(\bar{d}_\sigma,d_\sigma)$  
can be integrated out at the cost of the path summation over the 
  HS Ising spins $\{s\}$,  
\begin{equation}
Z[\eta]=\sum_{\{s\}}\prod_\sigma \det (-i G_\sigma^{-1}[\{ s \},\eta])  ,
\label{pathsum}
\end{equation}
with  the total Keldysh GF written in time-discretized 
($1\le k,l \le N$) form as
\begin{equation}
\left(G_{\sigma}^{-1}\right)_{kl}^{\alpha \beta}[\{ s \},\eta] 
= \left(G^{-1}_{0\sigma}\right)_{kl}^{\alpha \beta}
+i\eta\Sigma^{J,\alpha \beta}_{kl}
-i\delta_t \delta_{kl}\lambda_\alpha s^\alpha_k
\delta_{\alpha \beta},
\end{equation}
where $\alpha,\beta=\pm$ labels the Keldysh branches,
and the noninteracting GF is 
\begin{equation}\label{gij}
 G_{0\sigma,kl}=\int_{-\infty}^{\infty}\frac{d\omega}{2\pi}
e^{i\delta_t(k-l)\omega} G_{0\sigma}(\omega)  .
\end{equation}
Note that $G_{0\sigma}(t,t')$ depends only on time differences 
due to time-translational invariance of the noninteracting part. 
Moreover, the self-energy kernel stemming from the external 
source term follows as 
\begin{equation}
\Sigma^{J,\alpha\beta}_{kl}=\frac{e}{2 }
\left[ \gamma_{L,kl}^{\alpha\beta}- \gamma_{R,kl}^{\alpha\beta}\right]
\left[ \delta_{mk} \delta_{\alpha,+}-\delta_{ml} \delta_{\beta,+}\right],
\end{equation} 
where $\gamma_{p,kl}=\gamma_p(t_k-t_l)$ and the measurement time
is $t_m=m \delta_t$.

\section{The iterative path-integral scheme}
\label{sec:IPI}

\begin{figure}[t!]
\vspace{0.3cm}
\includegraphics[width=85mm]{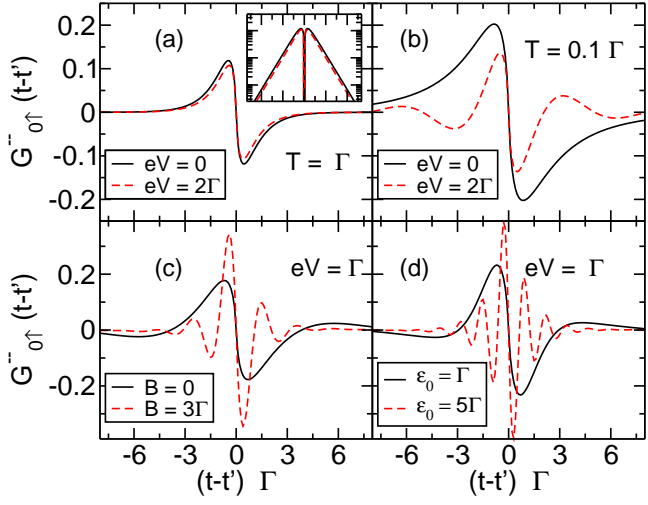}
\caption{ \label{corrfun} (Color online) Real part of the 
Keldysh GF component 
$G^{--}_{0\uparrow}(t-t')$ for 
(a) $\epsilon_0=0, B=0, T=\Gamma$, 
(b) $\epsilon_0=0, B=0, T=0.1\Gamma,$ 
(c) $\epsilon_0=0, T=0.1 \Gamma, eV=\Gamma, B=0$  and $B=3\Gamma$, 
(d) $B=0,  T=0.1 \Gamma, eV=\Gamma,  \epsilon_0=\Gamma$ and $
\epsilon_0=5\Gamma$. The inset in (a) shows the absolute values of the
 same data in log-linear representation, highlighting the exponential decrease. }
\end{figure}

Let us now exploit the property
(see, e.g., Ref.\ \cite{Weiss}) that each Keldysh 
component of $G_{0\sigma,kl}$ decays exponentially 
at long time differences ($|k-l|\to\infty$) for finite $T$, see Eq.\
(\ref{gij}).  We denote the corresponding  
time scale (correlation or memory time)
by  $\tau_c$.  In Fig.\ \ref{corrfun} we show typical examples of 
${\rm Re \,} G^{--}_{0\uparrow,kl}$ for different  bias 
voltages $V$. The exponential decrease with time 
is illustrated in the inset of Fig.\ \ref{corrfun}(a) where 
the absolute value $|{\rm Re \,} G^{--}_{0\uparrow,kl}|$ is again plotted for the same 
parameters, but in a log-linear representation. For large bias voltages
 and low enough temperatures, e.g., at  $V\gtrsim \Gamma$ and 
 $T\leq 0.2 \Gamma$, the decay is superposed by an oscillatory behavior, 
see Fig.\ \ref{corrfun}. Since the 
 lead-induced  correlation function decays as 
$\sim \cos[eV (t-t')/2]/\sinh[\pi T(t-t')]$, the  
respective correlations decay on a time scale 
given by $\tau_c^{-1}\sim \max(k_B T,eV)$. (The correlation function 
also has an additional $\epsilon_0$-dependence 
of the decay characteristics.) Thus, the exponential decay
 suggests to neglect lead-induced correlations beyond the 
correlation time $\tau_c$. This motivates an iterative scheme which
exactly takes into account the correlations within $\tau_c$, but
neglects them outside. The exponential decay has to be 
contrasted to the case $T=V=0$, where correlations die 
out only algebraically, and our approach is not applicable. 

Let us then face the remaining path sum in Eq.\ (\ref{pathsum}). 
In the discrete time representation, we denote
with $t_0=0<t_N=N\delta_t$ the initial and final time, and 
 $t_k=k\delta_t$, see Fig.\ \ref{Keldysh}. 
The discretized GF and self-energy kernels for given spin $\sigma$ 
 are then represented as matrices 
of dimension $2N\times 2N$. 
For explicit calculations, we arrange the matrix elements 
 related to Keldysh space (characterized by the Pauli matrices 
 $\tau$) and to  physical times $(t_k,t_l)$ as 
$\tau\otimes(k,l)$. 
In particular, the ordering of the matrix elements 
from left to right (and from top to bottom) 
represent increasing times. 
The lead-induced correlations thus decrease exponentially   
with growing distance from the diagonal of the matrix. 
For numerical convenience, we evaluate the generating  function in the 
equivalent form
\begin{equation}
Z[\eta]={\cal N}\sum_{\{s\}}\prod_\sigma \det D_\sigma[\{s\},\eta] ,
\label{pathcom}
\end{equation}
with $D_\sigma= G_{0\sigma} G_\sigma^{-1}$. Explicitly, this reads
\begin{equation}
D_{\sigma,kl}^{\alpha\beta}[\{ s \},\eta] =
 \delta_{\alpha \beta}- i \delta_t\lambda_\alpha  G_{0\sigma,kl}^{\alpha\beta} 
s^\alpha_k +i\eta\sum_{j,\alpha'} G_{0\sigma, kj}^{\alpha\alpha'}
\Sigma^{J,\alpha'\beta}_{jl} 
\label{dij}
\end{equation}
The normalization prefactor ${\cal N}$ does not affect 
physical observables, and is put to unity.  
The time local nature of the on-dot interaction is now present only  
in disguise, since the matrix product lets 
Ising spins appear line-wise. 
By construction, we have to sum over $2N$ auxiliary Ising spins, and 
the total number of possible spin configurations is $2^{2N}$. 

Next, we exploit the above-mentioned truncation  of the GF  by setting 
 $D_{kl}\equiv 0$ for $|k-l|\delta_t >\tau_c$, where 
 \begin{equation}
 \tau_c \equiv K \delta_t
 \end{equation} 
is the memory time, with $K$ the respective 
number of Trotter time slices. The GF matrices then have a $K$-band structure. 
Note that in the continuum limit, where $K= N, \delta_t\to 0$ and 
$N\to \infty$,  the approach is exact. 

To prepare the basis for the iterative scheme, we now transform the 
GF matrix to Schur's form, which then allows to calculate the determinant 
in a straightforward way.  
In general, a quadratic block matrix $D$ given as 
$D=\left(\begin{array}{cc}
a&b\\c&d \end{array}
\right)$,  with $a$ being invertible,  
can be represented in its Schur form, i.e.,  
after one step of Gaussian elimination. Hence, by 
multiplying from the left with 
a lower triangular matrix, 
\begin{displaymath}
\tilde{D}=LD=
\left(
\begin{array}{cc}
I_n&0\\
-ca^{-1}&I_m
\end{array}
\right)\left(
\begin{array}{cc}
a&b\\
c&d
\end{array}
\right)=
\left(
\begin{array}{cc}
a&b\\
0&d-ca^{-1}b
\end{array}
\right)\, ,
\end{displaymath}
where $I_{n(m)}$ represent unit matrices of dimension $n=\dim a  
\, \, (m=\dim d)$, and 
 $b,c$ do not have to be quadratic themselves. Clearly,
the determinant is invariant under this transformation. 
 The $(2,2)$ element in this block notation is often 
referred to as the Schur complement of the matrix $D$. 
This representation thus allows to write the determinant as
$\det(D)=\det(a)\det(d-ca^{-1}b).$
We can now establish the iterative scheme as
follows. We start from the
full GF in the matrix representation as defined in Eq.\ (\ref{dij}).
After the memory truncation, the $N\times N-$matrix (in  time
space) assumes a band structure represented as 
\begin{widetext}
\psset{xunit=1cm, yunit=1cm, linestyle=solid}
\begin{displaymath} 
  D \equiv  D_{(1,N_K)}
   = \left(
    \begin{array}{cccccc}
      \psframebox{D^{11}}&\psframebox{D^{12}}&0&0 & \dots & 0\\
      &&&&&\\
      \psframebox{D^{21}}&\psframebox{D^{22}}&\psframebox{D^{23}}&0&
      \dots& \vdots\\
      &&&&&\\
      0&\psframebox{D^{32}}&\psframebox{D^{33}}&\psframebox{D^{34}}&
      \dots &\vdots\\
      &&&&&\\
      0&0&\psframebox{D^{43}}&\psframebox{D^{44}}& \dots & 0 \\
      \vdots & \vdots &  \vdots & \vdots & \ddots &
      \psframebox{D^{N_K-1 N_K}}\\
      &&&&&\\
      0 & \dots & \dots & 0 & \psframebox{D^{N_K N_K-1}} &
      \psframebox{D^{N_K N_K}} \\
    \end{array}
  \right)
\end{displaymath}
where the single blocks are $K\times K-$block matrices defined as 
($l,l'=1, \dots, N_K$)
\begin{displaymath}
\psframebox{D^{ll'}}=\left(
\begin{array}{cccc}
D_{(l-1)K+1,(l'-1)K+1}&\dots&D_{(l-1)K+1,l'K}\\
\vdots&\ddots&\vdots\\
D_{lK,(l'-1)K+1}&\dots&D_{lK,l'K}
\end{array}
\right)\, .
\end{displaymath}
The number $N$ of Trotter slices is always chosen such
that $N_K\equiv N/K$ is integer. 
The elements $D_{kl}$ are given in Eq.\ (\ref{dij}), with their
dependence on the Ising spins $s_k^\pm$ kept implicit.   
Each $D_{kl}$ still has a $2\times 2-$Keldysh
structure, and a $2\times 2$ spin structure. 
Then, we rewrite the generating function (\ref{pathcom}) as 
\begin{eqnarray}
Z[\eta]&=&\sum_{s_1^\pm,\dots, s_N^\pm}\det\left\{D^{11}[s_1^\pm,\dots, s_K^\pm]\right\} 
\nonumber \\
& \times & \det\left\{D_{(2,N_K)}[s_{K+1}^\pm,\dots, s_N^\pm]
-D^{21}[s_{K+1}^\pm,\dots,s_{2K}^\pm] 
\left[ D^{11}[s_1^\pm,\dots,s_K^\pm]\right]^{-1}
D^{12}[s_{K+1}^\pm,\dots,s_{2K}^\pm]
\right\} \, ,
\label{zeta}
\end{eqnarray}
where the  $N_K-1\times N_K-1-$matrix $D_{(2,N_K)}$ is obtained from
$D_{(1,N_K)}$ by removing
the first line and the first column. 

In order to set up an iterative scheme, we use the following
observation: to be consistent with the
 truncation of the correlations after a memory time $K\delta_t$, 
 we have to neglect terms that 
directly couple Ising spins at time differences 
larger than the memory time. This is achieved by setting 
\begin{equation}
D^{l+2,l+1}D^{l+1,l} \left[D^{l,l}\right]^{-1} D^{l,l+1}  D^{l+1,l+2}
\to  0
\end{equation}
within the Schur complement in each further iteration step. 
Note that we do not neglect the full
Schur complement but only those parts which are generated in the
second-next iteration step. 
With this, we rewrite the generating function  as 
\begin{eqnarray} \label{zeta2}
Z[\eta] & = & \sum_{s_1^\pm,\dots, s_N^\pm} 
\det \left\{ D^{11}[s_1^\pm,\dots,s_K^\pm]\right\}
\prod_{l=1}^{N_K-1}\det \Big\{ D^{l+1,l+1}[s_{lK+1}^\pm,\dots,s_{(l+1)K}^\pm]
 \nonumber \\
 && -D^{l+1,l}[s_{lK+1}^\pm,\dots,s_{(l+1)K}^\pm]
\left[D^{l,l}[s_{(l-1)K+1}^\pm,\dots,s_{lK}^\pm]\right]^{-1}
D^{l,l+1}[s_{lK+1}^\pm,\dots,s_{(l+1)K}^\pm]
\Big\} \, . 
\end{eqnarray}
Then, one can exchange the sum and the product, and 
by reordering the sum over all Ising spins, one obtains
\begin{equation}
Z[\eta] = \sum_{s_{N-K+1}^\pm, \dots, s_N^\pm} Z_{N_K} 
[s_{N-K+1}^\pm, \dots, s_N^\pm] \, ,
\end{equation}
where $Z_{N_K}$ is the last element obtained from the 
iterative procedure defined by ($l=1, \dots, N_K-1$)
\begin{equation}
Z_{l+1}[s_{lK+1}^\pm,\dots,s_{(l+1)K}^\pm]=
\sum_{s_{(l-1)K+1}^\pm,\dots,s_{lK}^\pm}
\Lambda_{l}[s_{(l-1)K+1}^\pm,\dots, s_{lK}^\pm,s_{lK+1}^\pm,\dots,s_{(l+1)K}^\pm]
Z_{l}[s_{(l-1)K+1}^\pm,\dots,s_{lK}^\pm] \, .
\label{zit}
\end{equation}
The  {\it propagating tensor} $\Lambda_{l}$  can be read off
from Eq.\ (\ref{zeta2}) as
\begin{eqnarray}
\Lambda_{l}&=&\det \Big\{ D^{l+1,l+1}[s_{lK+1}^\pm,\dots,s_{(l+1)K}^\pm]
 \nonumber \\
 && -D^{l+1,l}[s_{lK+1}^\pm,\dots,s_{(l+1)K}^\pm]
\left[D^{l,l}[s_{(l-1)K+1}^\pm,\dots,s_{lK}^\pm]\right]^{-1}
D^{l,l+1}[s_{lK+1}^\pm,\dots,s_{(l+1)K}^\pm]
\Big\}\, .
\label{lambda}
\end{eqnarray}
\end{widetext}
Note that we use here the notion of {\em tensor\/} in the sense of a
multi-dimensional array and not in the strict mathematical sense
defined by the transformation properties of this object, see also 
Ref.\ \cite{Makri}. The iteration starts with 
 $Z_{1}[s_{1}^\pm,\dots,s_{K}^\pm]=
 \det \left\{D^{11}[s_1^\pm,\dots,s_K^\pm]\right\}$. 

The current is numerically obtained by evaluating 
Eq.\ (\ref{gencurrent})  
for a small but fixed value of $\eta$; we have taken 
$\eta=0.001$ for all results shown below. By this, we obtain 
the full time-dependent current 
$I(t_m)$ as a function of the measurement time $t_m$ 
($0\le t_m \le N\delta_t$). At short times, this
shows a transient oscillatory or relaxation behavior,
which then reaches a plateau value from which we read off
the steady-state current $I$. 

\section{Convergence and extrapolation procedure}
\label{sec:extrapol}

\begin{figure}[t!]
\vspace{0.5cm}
\includegraphics[width=85mm]{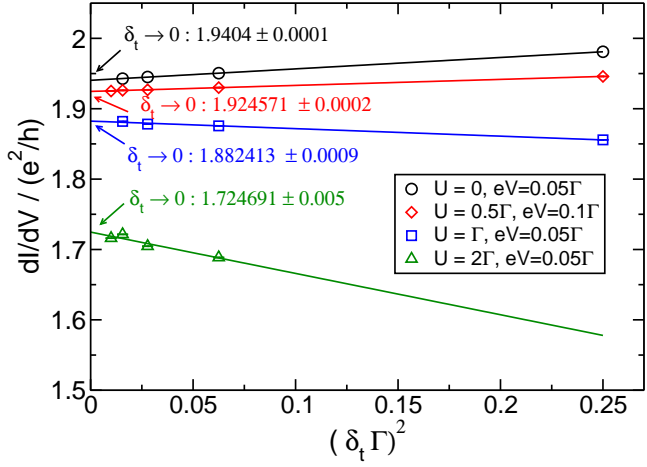}
\caption{ \label{tautrott} (Color online) Quadratic dependence of the 
Trotter error as obtained in the convergence procedure for 
$\delta t\to 0$ for a fixed memory time $\tau_c=0.5/ \Gamma$. Parameters are 
  $\epsilon_0=0, B=0, T=0.1 \Gamma$.  
 The extrapolated values are
given in the figure. }
\end{figure}

\begin{figure}[t!]
\vspace{0.2cm}
\includegraphics[width=85mm]{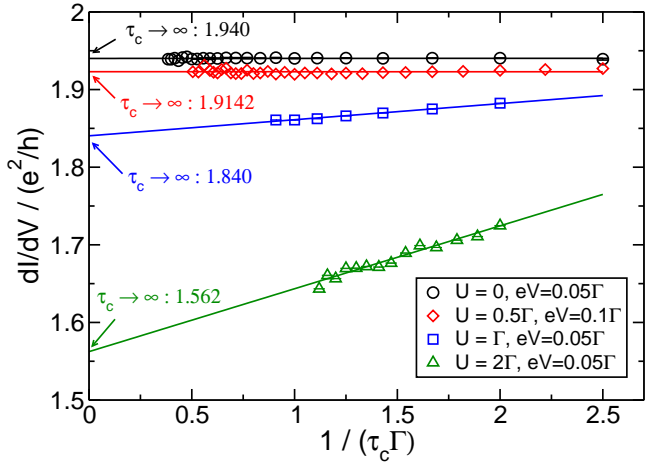}
\caption{ \label{taumem} (Color online) Dependence of the differential
conductance  $dI(\tau_c)/dV$ 
on the inverse 
memory time $1/\tau_c$ for different values of $U$ after the Trotter
error has been eliminated. 
Parameters are $\epsilon_0=B=0,T=0.1\Gamma$. 
Solid lines correspond to a linear fit. 
The extrapolated values for $dI/dV=dI(1/\tau_c \to 0)/dV$ 
are given in the figure.}
\end{figure}

In order to render the scheme exact, we have to eliminate the two
systematic errors which are still present up to this point, namely, 
 (i) the Trotter error due to finite time
discretization $\delta_t=t/N$, and (ii) the memory error due to a
finite memory time $\tau_c = K \delta_t$. The scheme becomes exact by
construction in the limit $K\to \infty$ and $\delta_t \to 0$. 
To perform this limit in a straightforward way is not
possible due to the exponential dependence of the array sizes on $K$.
However, unless temperature is very low, we can eliminate both errors from the
numerical data in the following systematic way:  
(i) We choose a fixed
  time discretization $\delta_t$ and a 
  memory time $\tau_c$. A reasonable estimate for $\tau_c$ is 
 the minimum of $1/|eV|$ and $1/T$ (see
above). With that, we calculate the current $I(\delta_t, \tau_c)$, 
and, if desired, the differential conductance $dI(\delta_t, \tau_c)/dV$ 
(the derivative is performed numerically for
a small $\Delta V=0.01 \Gamma$). The calculation is
then repeated for
different choices of $\delta_t$ and $\tau_c$.  
(ii) Next, the Trotter error can be eliminated by exploiting 
the fact that it  vanishes quadratically for  
$\delta_t\to 0$ \cite{HirschFye,Hirsch,Fye}. For  a fixed  memory 
time $\tau_c$, we can thus extrapolate  and obtain 
$dI(\tau_c)/dV=dI(\delta_t \to 0, \tau_c)/dV$, which still
 depends on the finite memory time $\tau_c$. The quadratic dependence on
 $\delta_t$ is illustrated  in Fig.\  \ref{tautrott} for different
 values of $U$. Note that each line corresponds to the same fixed
 memory time $\tau_c=0.5/\Gamma$.   
 (iii) In a last step, we eliminate the memory error by  
 extrapolating  for $1/\tau_c \to 0$, and obtain
 the final numerically exact value $dI/dV=dI(\tau_c \to \infty)/dV$. 
For the  dependence on $1/\tau_c$,  we empirically
 find a regular and systematic behavior as shown in Fig.\
 \ref{taumem}. The $\tau_c \to \infty$ value is approached with
 corrections of the order of $1/\tau_c$, see Fig.\
 \ref{taumem}. However, it should be stressed that temperature and voltage 
affect the convergence properties in different ways, as is already
clear from Fig.~\ref{corrfun}. Importantly enough, when $T$ and $V$ are
such that the extrapolation scheme described above converges, numerical
exactness is warranted.   
 
We have implemented the  iterative scheme together with the convergence
procedure on standard Xeon 2GHz machines. 
Computations are then only possible for $K\leq 7$ due to 
the limited memory (RAM) resources available.
Typical running times for the shown simulation data are 
approximately $15$ hours for $K=5$. With a second, parallelized version of the 
code, 
we have been able to take into account up to $2^{14}$ summands in 
Eq.\ (\ref{zit}), corresponding to $K=7$, within passable running times of a few days. 
 
\section{Results} 
\label{sec:Results}

\begin{figure}[t!]
\vspace{0.2cm}
\includegraphics[width=85mm]{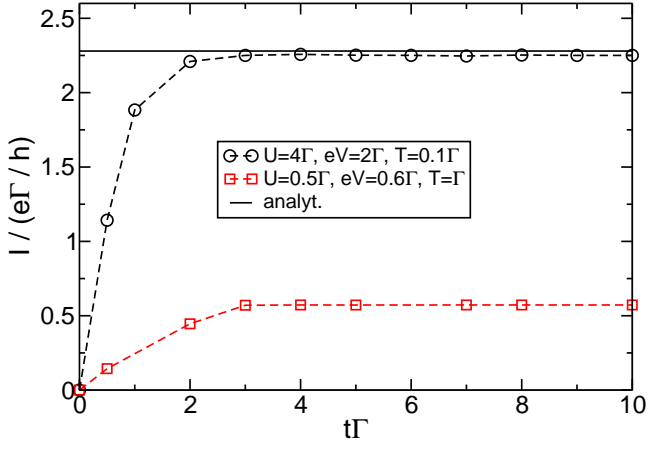}
\caption{ \label{ioft} (Color online) Time-dependent current $I(t)$  for 
two representative parameter sets: (1) 
$U=4 \Gamma, eV=2 \Gamma, T=0.1 \Gamma$ (black circles)
and (2) $U=0.5 \Gamma, eV=0.6 \Gamma, T=\Gamma$ (red squares).
Solid line: analytical result for stationary $I$ from 
nonequilibrium Kondo theory applied to parameter set (1)
 \cite{Kaminski00,Rosch05}.  Symbols are ISPI results, dashed 
 lines are guides to the eye only,
 and $\epsilon_0=B=0$. }
\end{figure}

Next, we discuss the results obtained by the application 
of the iterative procedure to
the Anderson model. As pointed out before, see Sec.\ \ref{sec:nonin}, 
we consider the symmetric case, $\Gamma_R=\Gamma_L=\Gamma/2$, with
$\mu_{L/R}=\pm eV/2$. In what 
follows, we measure energies in units of $\Gamma$. Unless noted otherwise, 
all error bars for the shown data points, 
which are due to the Trotter and memory extrapolation scheme, are of 
the order of the symbol sizes in the figures.  
Our scheme yields the full time-dependent current $I(t)$, including
transients as well as the asymptotic steady-state value. 
Fig.\ \ref{ioft} shows typical results for the current $I(t)$,  for 
two parameter sets, namely (1) 
$U=4 \Gamma, eV=2 \Gamma, T=0.1 \Gamma$ (black circles),
and (2) $U=0.5 \Gamma, eV=0.6 \Gamma, T=\Gamma$ (red squares).
The first set is of interest in the context of the nonequilibrium 
Kondo effect, where the Kondo temperature (for $\epsilon_0=0$) 
is $T_K=\sqrt{\Gamma U/2}\exp(-\pi U / 8\Gamma)$;
for parameter set (1), this yields $T_K=0.29 \Gamma$.
For $eV\gg T_K$, analytical results for the steady-state
current are available \cite{Kaminski00,Rosch05},  
shown as solid line in Fig.\ \ref{ioft}. This indicates
that the ISPI method is capable of approaching
the nonequilibrium Kondo problem.  For both parameter sets, 
and for many others not shown here, we observe that
after a transient relaxation behavior, $I(t)$ settles at a plateau value,
which then defines the stationary current discussed in the following.

\subsection{Validation of the algorithm: 
comparison with exact and perturbative results}

\begin{figure}[t!]
\vspace{0.3cm}
\includegraphics[width=85mm]{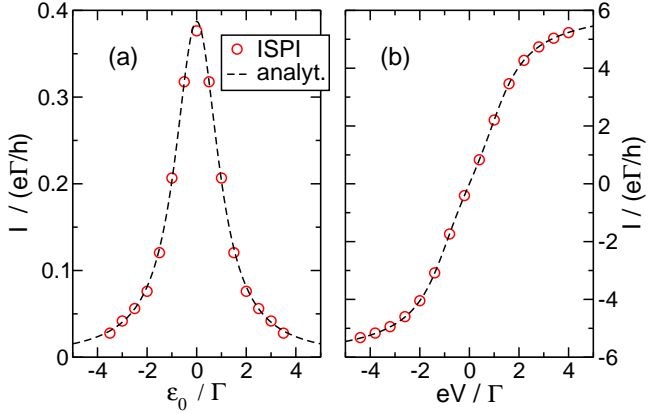}
\caption{ \label{Ueq0} (Color online) Stationary current as a function of 
$\epsilon_0$ (a) and of the bias voltage $eV$ (b) for the noninteracting case $U=0$. 
Shown are the numerical (circles) and the exact analytical (dashed curve) results for (a)  
$eV=0.2\Gamma$, and (b) $\epsilon_0=0$. In both cases, $T=0.1\Gamma, B=0$. 
}
\end{figure}

As a simple warm-up check, let us briefly compare our numerical results to 
the exact result for $U=0$. Figure \ref{Ueq0}(a)  
shows the stationary current (obtained already at $t_m=10 / \Gamma$) as a function of $\epsilon_0$ 
for $T=0.1\Gamma$ and $eV=0.2 \Gamma$. The $I$-$V$ characteristics is shown 
in Fig.\ \ref{Ueq0}(b) for $T=0.1 \Gamma$ and $\epsilon_0=0$.  
Both cases illustrate that the numerical result
coincides with the exact one.  Other parameter 
sets have been checked as well, and agree with the well-known
$U=0$ analytical solution.  

\begin{figure}[t!]
\vspace{0.2cm}

\includegraphics[width=85mm]{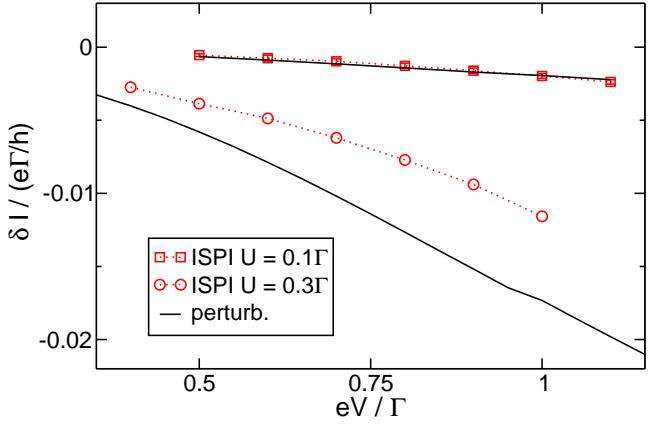}
\caption{ \label{pert_u01} (Color online) 
Interaction corrections $\delta I(V)=I(U)-I(U=0)$ from  
the numerical ISPI  approach (red symbols) and from a second-order 
perturbative calculation (black solid lines),
 for $U=0.1 \Gamma$ (squares) and $U=0.3 \Gamma$ (circles). 
Parameters are $\epsilon_0=B=0, T=0.1\Gamma$, and 
 dotted lines are guides to the eye only. }
\end{figure}

Second, in order to validate the reliability of
our code for finite $U$, we compare the numerical results 
to a perturbative calculation, 
where the interaction self-energy is computed 
up to second order in $U$ \cite{Luca}. In order to respect current conservation,
this calculation is possible only 
at the electron-hole symmetric point, $\epsilon_0=B=0$ \cite{Fujii}. 
First-order terms give then no contribution, and the self-energy corresponds
to just one diagram \cite{Luca}. 
For a detailed comparison, we plot 
mostly the {\sl interaction corrections},  $\delta A\equiv A(U)-A(U=0)$, 
with $A$ being the current $I$,  the linear conductance $G$,  or 
the nonlinear conductance $dI/dV$, respectively.    
Figure \ref{pert_u01} shows the results for $\delta I$ as a function 
of the bias voltage for $U=0.1 \Gamma$ and 
$U=0.3 \Gamma$. For $U=0.1 \Gamma$,  we perfectly recover the perturbative 
results, which confirms the reliability of our code even in the regime 
of nonlinear transport. Clearly,  
 the  corrections are small and negative, 
which can be rationalized in terms of Coulomb blockade physics, as  
transport is suppressed by a finite interaction on the dot.

\begin{figure}[t!]
\vspace{0.2cm}
\includegraphics[width=85mm]{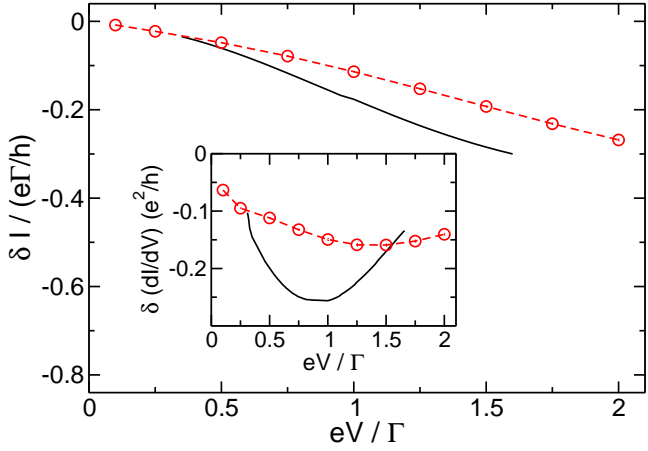}
\caption{ \label{pertu1} (Color online) 
Interaction correction $\delta I$ for the nonlinear current 
 for $U=\Gamma$,
 comparing ISPI (red symbols) and $U^2$ perturbation theory 
 (black solid curve). The inset shows the corresponding 
 interaction corrections $\delta (dI/dV)$ to
 the differential conductance. 
 Other parameters are as in Fig.\ \ref{pert_u01}. 
 Dashed lines are guides to the eye
only.}
\end{figure}

For   $U=0.3\Gamma$, the current decreases even more, 
and the deviations between the 
ISPI and perturbative results are also larger.  
The relative deviation for $U=0.3\Gamma$ is already $\approx 30-35 \%$, 
illustrating that perturbation theory is already of limited accuracy in this 
regime. Although it well reproduces the overall tendency, there is 
significant quantitative disagreement. 
The differences are even more pronounced  for $U=\Gamma$, 
as shown in Fig.\ \ref{pertu1} for $\delta I$ (main) 
 and for $\delta(dI/dV)$ (inset).
Here, second-order perturbation theory does not even reproduce qualitative 
features.

\subsection{Comparison with  master equation approach}

\begin{figure}[t!]
\vspace{0.2cm}
\includegraphics[width=85mm]{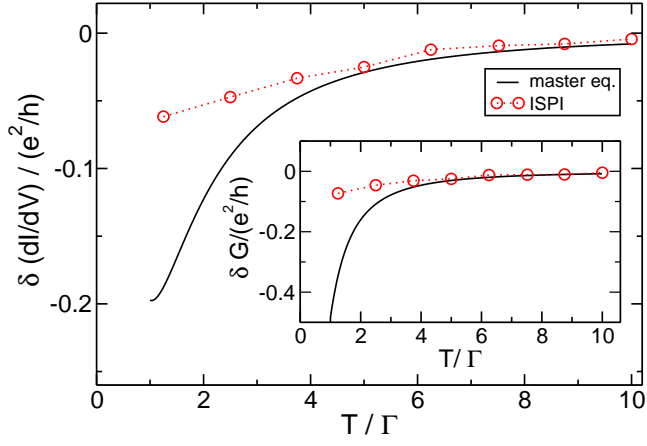}
\caption{ \label{masterlinnonlin} (Color online) 
Comparison of the ISPI method (red symbols) and the master equation 
approach (solid lines) 
 for $U=\Gamma$. Main: Corrections $\delta (dI/dV)$ of the nonlinear 
 conductance as a function of temperature for 
 $eV=3 \Gamma$. Inset: Same for the linear conductance $\delta G$.  
 For the remaining parameters, see Fig.\ \ref{pert_u01}.}
\end{figure}

Next, we compare our numerical approach with the outcome of  
a standard master equation calculation \cite{Flensberg}. 
The master equation is expected to
yield reliable results in the incoherent (sequential) tunneling regime,
 $T \gg \Gamma$, where a description in terms of occupation probabilities for
the isolated many-body dot states is appropriate. 
The transport dynamics is then described by a  
rate equation for the populations, where the time-dependent rates 
are obtained from lowest-order perturbation theory in $\Gamma$ \cite{Flensberg}.  
The results are shown for $U=\Gamma$ in Fig.\ \ref{masterlinnonlin}, 
both for the nonlinear and the linear differential conductance
interaction corrections. As temperature is lowered, 
interaction effects become more important, as seen from the exact
ISPI results. 
 This corresponds to the emergence of coherence effects for $T<\Gamma$,
which are clearly not captured by the master equation in the sequential 
tunneling approximation.
  However, for $T\agt 4\Gamma$,
interaction corrections are washed out, and the master equation 
becomes essentially exact, cf.~Fig.~\ref{masterlinnonlin}.  
Similarly, from our ISPI results, we observe that 
interaction corrections are 
suppressed by an increasing bias voltage as well.
To give numbers, for $T=1.25\Gamma$, we find in the linear regime
 $\delta G=-0.073e^2/h$, 
whereas $\delta (dI/dV)=-0.062e^2/h$ for $eV=3\Gamma$. 

\subsection{Small bias: $eV \ll \Gamma$}
\label{sec:linear}

\begin{figure}[t!]
\vspace{0.2cm}
\includegraphics[width=85mm]{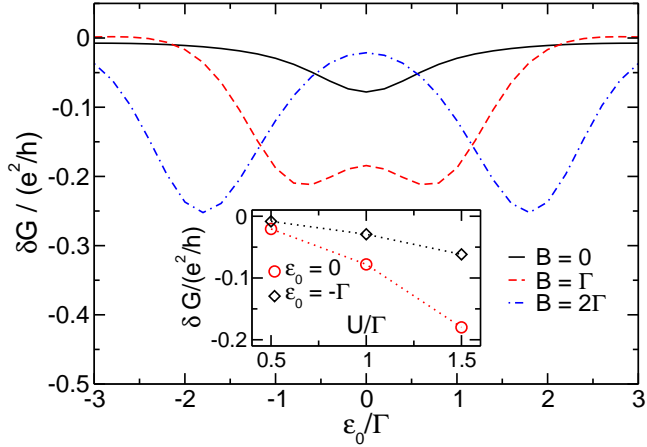}
\caption{ \label{Gvseps01} (Color online) 
Interaction correction $\delta G$ as a 
function of $\epsilon_0$, for $U=\Gamma$, different $B$, and $T=0.1\Gamma$.
Inset: Dependence of $\delta G$ on $U$ for $B=0$, on resonance ($\epsilon_0=0$, red
circles) and away from resonance ($\epsilon_0=-\Gamma$, black diamonds).} 
\end{figure}

For sufficiently small bias voltage, 
the current is linear in $V$, and we can focus 
 on the linear conductance and its interaction correction $\delta G$.  
Figure \ref{Gvseps01} shows $\delta G$ as a function of  $\epsilon_0$
for different magnetic fields $B$, 
taking $U=\Gamma$ and $T=0.1$ (to be specific, we have chosen $eV=0.05\Gamma$). 
For $B=0$, two spin-degenerate transport channels contribute,
and a single resonant-tunneling peak at $\epsilon_0=0$ results. 
The interaction corrections are most pronounced on resonance, $\epsilon_0=0$. 
For $B\ne 0$, the spin-dependent channels are split
 by $\Delta \epsilon=2B$, resulting in a double-peak
structure. The spin-resolved levels are now 
positioned at $\epsilon_0\pm B$ due to the Zeeman splitting.
Again the interaction corrections are largest on resonance, 
and the double-peak
structure of $G(\epsilon_0)$ is transferred to $\delta G(\epsilon_0)$, 
cf.~Fig.\  \ref{Gvseps01}.  We find no evidence for an interaction-induced 
broadening of the resonant-tunneling peak compared to the noninteracting case. 
The width of the Lorentzian peak profile for $B=0$ is determined by 
$\Gamma$ at sufficiently low $T$, and broadens as $T$ increases. 
Here, the double-peak structure, with two clearly separated
peaks for  finite $B$, is not yet fully developed. The two peaks largely 
overlap, and the distance of the peaks  is below the expected  
$\Delta \epsilon =2B$, since tunneling considerably 
broadens  the dot levels. We note however, that for larger $B$, 
the correct peak spacing of $\Delta \epsilon = 2B$ is reproduced 
(data not shown). 

In order to illustrate the role of the interaction $U$,
we show the dependence of $\delta G$ on $U$ at $\epsilon_0=0$
 in the inset of Fig.\  \ref{Gvseps01}. 
For $U=0$, both levels contribute $e^2/h$ to the conductance at 
low temperatures, while 
 for finite $U$ these contributions are reduced. The reduction 
 increases with growing $U$,
 qualitatively consistent with  previous results. The interaction 
 corrections are smaller away from resonance, see inset 
of Fig.\  \ref{Gvseps01}, but they also grow with increasing $U$. 

\begin{figure}[t!]
\vspace{0.2cm}
\includegraphics[width=85mm]{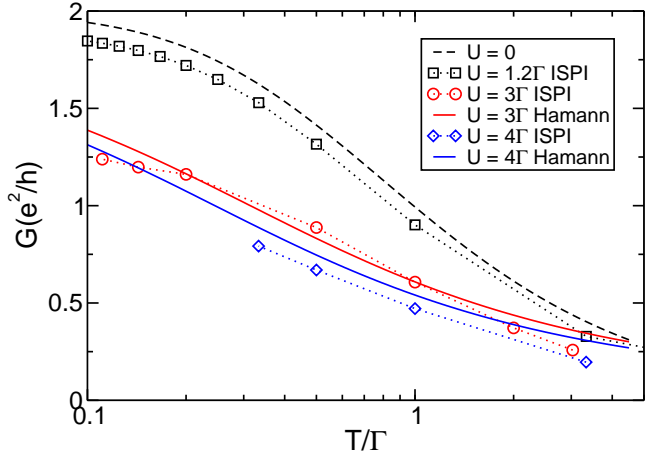}
\caption{ \label{lingoft} (Color online) Linear 
 conductance $G$ vs  
temperature $T$,  for 
$U= 0$ (dashed line) and $U=1.2, 3, 4\Gamma$ ($\epsilon_0= B=0$). 
Symbols denote the ISPI results while the
solid lines are those of Eq.\ (\ref{hamann}).  
} 
\end{figure}

Next, we address the temperature dependence of the linear conductance 
(numerically evaluated for $eV=0.05 \Gamma$). In Fig.\ 
\ref{lingoft}, we show $G(T)$ for different values of $U$ 
(up to $U=4 \Gamma$) at $\epsilon=B=0$. For $U=1.2 \Gamma$, the 
deviation from the $U=0$-result is small in  the 
considered temperature range. 
For larger $U$, deviations become more pronounced at low temperatures
where interaction becomes increasingly relevant.  
Up to present, we have obtained converged results in the regime 
of {\em small\/} bias voltages for interaction strengths 
$U\leq 4 \Gamma$ for temperatures above or close to the Kondo temperature,  
$T\gtrsim T_K$. 
The corresponding Kondo temperatures are (see above) $T_K=0.38\Gamma$ for 
$U=3\Gamma$ and $T_K=0.293 \Gamma$ for $U=4 \Gamma$. 
In the regime $T_K \lesssim T \lesssim 10 T_K$, we can compare our results 
to the result of Hamann \cite{Costi2,Hamann},
\begin{equation}\label{hamann}
G(T)=\frac{e^2}{h} 
\left(1-\frac{\ln(T/T_{KH})}{[\ln^2(T/T_{KH})+3\pi^2/4]^{1/2}}\right) \, ,
\end{equation}
for the linear conductance, where $T_{KH}=T_K/1.2$, see Fig.\ \ref{lingoft} 
(solid lines). 
In Ref.\ \cite{Costi2}, it has been shown that the results of 
 the numerical RG coincide with those of Eq.\ (\ref{hamann}) in this regime. 
 Fig.\ \ref{lingoft} illustrates that 
 the agreement between the two approaches is satisfactory and 
 shows that the ISPI provides reliable results 
 in the linear regime above or close to the Kondo
 temperature. As already mentioned, convergence is problematic 
in the {\em linear\/} regime for temperatures lower than the Kondo 
temperature for larger values of $U$. This implies that 
 the equlibrium Kondo regime is difficult 
to explore using the ISPI approach. However, the situation is 
more favorable for large bias voltages, where short to intermediate 
memory times
are sufficient, and we have achieved convergence up to $U=4\Gamma$, 
see Fig.~\ref{ioft} and the next subsection.

\subsection{Large bias: $eV \ge \Gamma$}
\label{sec:noneq-res}

\begin{figure}[t!]
\vspace{0.2cm}
\includegraphics[width=85mm]{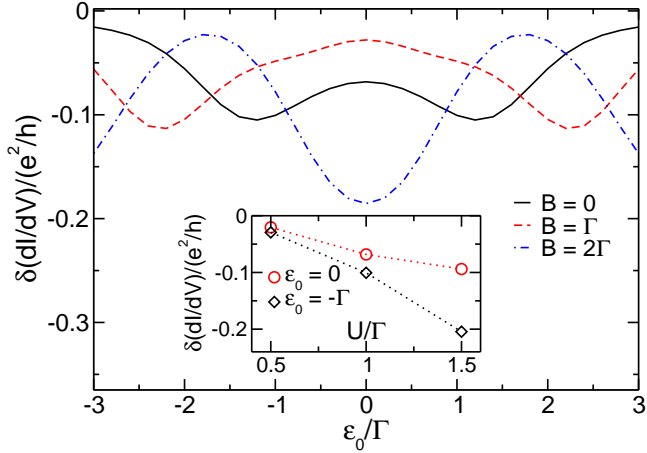}
\caption{ \label{Gvseps04} (Color online)
Same as Fig.~\ref{Gvseps01} but for the nonlinear conductance, taken at  
 $eV=3 \Gamma$. }
\end{figure}

Let us then turn to nonequilibrium transport  at voltages 
  $eV\agt \Gamma$.
Here,  the transport window is $\sim eV$, and a 
double-peak structure for $dI/dV$ emerges even for $B=0$,
with  distance $eV$ between the peaks. As interaction corrections 
to the nonlinear conductance 
are largest when a dot level is in resonance 
with one of the chemical potentials of the leads,
 the double-peak structure is again transferred to 
$\delta(dI/dV)$. This can be observed in Fig.\ 
\ref{Gvseps04}, where we show results for $eV=3\Gamma$ but
otherwise the same parameters as in Fig.\ \ref{Gvseps01}.
 For a finite magnetic field, each
peak of the double-peak structure itself experiences an additional 
 Zeeman splitting, resulting in an overall four-peak structure. 
 For $B=\Gamma$ and the
 shown values of $U$, 
 the two innermost peaks (closest to $\epsilon_0=0$) overlap
 so strongly that they effectively form a single peak at $\epsilon_0=0$ 
 again. The two outermost peaks are due to the combination of the 
 finite magnetic field and the bias voltage. 

Increasing the on-dot interaction $U$ leads to a reduction of
the differential conductance peaks compared to the noninteracting
case, i.e., the interaction corrections are again largest when the level 
energy matches the chemical potential in the leads. This is 
observed in the inset of Fig.\ \ref{Gvseps04}, for $B=0$ and 
$\epsilon_0=-\Gamma$, which is close to the peak maximum. Away from resonance, 
the interaction corrections are reduced. 
Note that the four-peak structure is already present in the
noninteracting case (with $B\ne 0$) 
and, hence, is not  modified qualitatively by $U$. 

In the regime $eV\gg T_K$, 
 we can compare our results to the perturbative RG result
 \cite{Kaminski00,Rosch05} 
 \[
I(V)=\frac{3\pi^2 e^2 V}{8h} \ln^{-2}(eV/\tilde{T}_K) \, ,
\] 
with $\tilde{T}_K=2 T_K/\sqrt{\pi}$. Fig.\ \ref{ioft} 
 shows the result for the stationary current for  
 $U=4\Gamma$ (where $T_K=0.29 \Gamma$) for $eV=2\Gamma$. 
 The perturbative current amounts to $I_{\infty}=2.28 e\Gamma/h$, 
while the ISPI value is $I_{ISPI}=2.25 e\Gamma/h$.  
 The quite satisfactory agreement suggests that ISPI is
indeed a reliable new method that holds promise for 
reaching the nonequilibrium Kondo regime. 
A detailed study of the nonequilibrium Kondo effect using ISPI will
be given elsewhere.

\begin{figure}[t!]
\vspace{0.2cm}
\includegraphics[width=85mm]{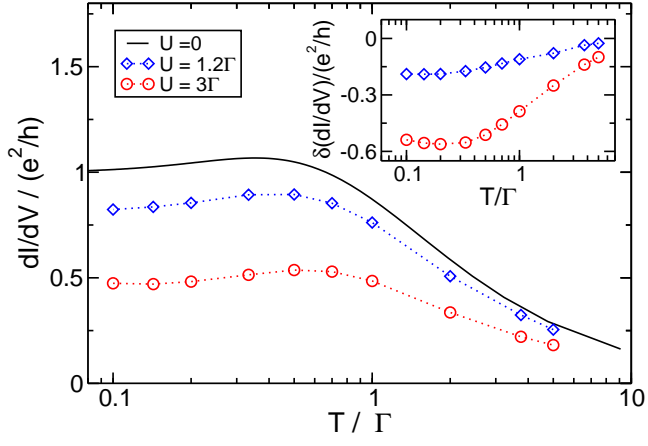}
\caption{ \label{GofT} (Color online) Nonlinear differential
 conductance $dI/dV$ vs  
temperature $T$,  for $eV=2\Gamma,\epsilon_0= B=0$, and 
$U=0, 1.2, 3\Gamma$.  
Inset: corresponding interaction corrections $\delta(dI/dV)$. 
(The Kondo temperature for $U=3\Gamma$ is
$T_K=0.38 \Gamma$.) 
} 
\end{figure}

Finally, we address the temperature dependence of 
the nonlinear conductance $dI/dV$. 
ISPI results for $eV=2 \Gamma, \epsilon_0=B=0$ are shown in 
 Fig.\ \ref{GofT}. Again, as in the linear regime, 
 the conductance increases  with lower temperatures,  
and finally saturates, e.g.,  at $dI/dV=e^2/h$ for $U=0$ and $eV=2\Gamma$. 
Clearly the conductance decreases when the bias voltage is 
raised. 
Increasing $U$ renders this suppression yet more pronounced, see also 
inset of Fig.\ \ref{GofT} for the corresponding corrections. 
At high temperatures,  thermal fluctuations wash out the interaction 
effects, and the interaction corrections die out. 

\section{Discussion and conclusions}
\label{sec:Conclusion} 

In summary, we have introduced a scheme for the iterative summation 
of real-time path integrals (ISPI), and applied it to
 a prototypical problem of quantum transport 
through an interacting quantum dot coupled to 
metallic leads held at different chemical  
potentials. After integrating out the 
leads, a time-nonlocal Keldysh self-energy 
arises. Exploiting the 
exponential decay of the time correlations at finite temperature 
allows to introduce a memory time $\tau_c$ 
beyond which the correlations can be truncated. 
Within $\tau_c$,  correlations are fully taken into account 
in the corresponding path integral for the current generating function. Then, 
through a discrete Hubbard-Stratonovich transformation, interactions can
be transferred to an auxiliary spin field, and 
an iterative summation scheme has been constructed
to calculate the transport current. The remaining systematic errors due to the finite time
discretization and the finite memory time $\tau_c$ 
are then eliminated by a refined Hirsch-Fye-type 
extrapolation scheme, rendering the ISPI
numerically exact.  From this construction, it is clear that the calculation is reliable
for temperatures above a certain interaction-dependent temperature scale. The latter depends also on the computational power available, and vanishes in the absence of interactions.

The general scheme has been applied to the canonical example of an Anderson dot with
interaction $U$, for which many
features of the transport characteristics are well understood. 
This allows to carefully and systematically check
the algorithm. In the regime of linear transport, we have recovered results 
from second-order perturbation theory in $U$ in the limit of very
small interaction strength, but 
found significant deviations already for
small-to-intermediate values of $U$. In the incoherent sequential regime, 
we recover results from a master
equation approach. Taking $U/\Gamma=4$,
we have furthermore reproduced the behavior of 
 the linear conductance above the Kondo temperature and found 
 satisfactory agreement with numerical RG results. 
 In addition, we have investigated the regime of 
correlated nonlinear transport, where, in our opinion, the 
presented method is most valuable.  We have checked that for large voltage,
known results on the nonequilibrium Kondo effect are reproduced as well.

Up to now, we have achieved converged results for small-to-intermediate 
$U$ even at rather low temperatures and small bias voltages. 
However, the strong-coupling regime of the 
Kondo effect has not yet been fully captured, as the required 
memory times become quite large,
and, at the same time, small $\delta_t$ are necessary because of the large $U$.  This combined
requirement makes it difficult to reach convergence.
Nevertheless, the {\em nonequilibrium\/} Kondo regime, representing 
an intermediate-to-weak coupling
 situation, seems tractable  by the ISPI scheme.
The applicability and accuracy of ISPI has been demonstrated for bias voltages  
larger than the Kondo temperature. 
We will present a detailed study of this interesting regime elsewhere.

Our approach is, in fact, similar in spirit to the
well-established concept of the
quasi-adiabatic path integral (QUAPI) scheme, introduced by Makri and Makarov 
\cite{Makri} in its iterative version. This method has been  developed to describe the dynamics 
of a quantum system coupled to a bath of harmonic oscillators held at equilibrium, 
in order to obtain exact results 
for quantum decoherence and dissipation,  see also Refs.~\cite{Eckel,Thorwart1}.
For a dipole-type system-bath coupling, as it occurs, e.g., in the spin-boson model \cite{Weiss}, 
the bath-induced correlations are encoded in the 
Feynman-Vernon influence functional \cite{Feynman}, which is 
solely a functional of a single {\em system\/}  operator, say $\hat{x}$, which couples the system to the
bath. The very existence
of this functional allows to perform a basis rotation to the
eigenbasis of $\hat{x}$ (called the discrete variable representation,
DVR). Then, the influence functional can be evaluated at the eigenvalues of $\hat{x}$ during
the numerical calculation of the real-time path integral over $\hat{x}(t)$. Moreover, within the QUAPI scheme, the  influence 
functional also generates correlations which are non-local in time, but which also decay exponentially.
This allows to truncate them beyond a  memory-time $\tau_{mem}$, 
which coincides conceptually with our correlation time $\tau_c$. 
Then, in practice, numerical convergence has to be achieved with respect to 
the memory time \cite{Eckel}. Afterwards, the only remaining 
Trotter error can be completely eliminated by extrapolation \cite{Eckel}. 
In the present case of nonequilibrium transport, 
several fundamental differences occur, and, in fact, only the
strategy of memory truncation can be taken over. The major difference 
 is that there exists no simple Feynman-Vernon influence functional.
While the lead fermions can be integrated out, 
see Eq.\ (\ref{Seff}), the corresponding Grassmann variables for the 
dot cannot be transferred to a practical computational scheme
 along the lines of the QUAPI method. 
Instead, here we integrate out {\em all\/} fermions, 
 at the expense of introducing auxiliary Ising spins
 via the HS transformation. Then, ISPI 
performs a summation over these Ising
spins. Another difference is the form of the
tunnel coupling in the Hamiltonian. As  two non-commuting system operators
$d_\sigma$ and  $d_\sigma^\dagger$ occur, no analogue of DVR can be established here.

Finally,  we note that we have chosen the Anderson model as a simple 
but non-trivial toy model for quantum transport in order 
to establish and test the numerical algorithm.
We believe that this approach, or modifications thereof, will be
of importance in numerical calculations of quantum transport properties.
Compared to other approaches, it has several advantages (e.g., numerical
exactness, direct nonequilibrium formulation, no sign problem), 
but is, on the other hand, also computationally more costly than most
other techniques, especially for strong correlations and/or low energy
scales (temperature, voltage). In any case, it
 goes without saying that
other interesting models for quantum transport exist, and
future work will be devoted to apply ISPI to those.

\acknowledgments
We acknowledge financial support within the DFG 
priority program 1243 ``Quantum transport at the molecular scale'' and by the 
SFB Transregio 12. 
We thank L. Dell'Anna for providing the results of the 
nonequilibrium perturbation theory, and for discussions. 
We also thank K. Flensberg, A.-P. Jauho and J. Paaske for 
useful discussions, and for 
the kind hospitality at the Niels-Bohr Institute of the 
University of Copenhagen (M.T.), where  
parts of this work have been completed. 
Computational time from 
the ZIM at the Heinrich-Heine-Universit\"at D\"usseldorf is acknowledged.

\newpage

\begin{thebibliography}{99}

\bibitem{Heinzel}
T.\ Heinzel, {\em Mesoscopic Electronics in Solid State 
Nanostructures\/}, 2nd ed.\  (VCH-Wiley, Berlin, 2006). 

\bibitem{molel}
{\em Introducing Molecular Electronics\/},
G. Cuniberti, G. Fagas, K. Richter (eds.), Lecture Notes in Physics, 
(Springer, Heidelberg, 2005).

\bibitem{Nitzan07}
M. Galperin, M.A. Ratner, and A. Nitzan, J. Phys. Cond.
Matt. {\bf 19}, 103201 (2007). 

\bibitem{Weber02}
J.\ Reichert, R.\ Ochs, D.\ Beckmann, H.\ B.\ Weber, M.\ Mayor, and 
H.\ v.\  L\"ohneysen, Phys. Rev. Lett. {\bf 88}, 176804 (2002). 

\bibitem{Ruitenbeek97}
R.\ Smit, Y.\ Noat, C.\ Untiedt, N.\ D.\ Lang, M.\ C.\ 
van Hemert, and J.\ M.\ Ruitenbeek, Nature {\bf 419}, 906 (2002). 

\bibitem{Scheer04}
T. B\"ohler, J.\ Grebing, A.\ Mayer-Gindner, H.\ v.\ L\"ohneysen, 
and E.\ Scheer, Nanotechnology {\bf 15}, 465 (2004).

\bibitem{Park} 
H. Park, J. Park, A.K.L. Kim, E.H. Anderson, 
A.P. Alivisatos, and P.L. McEuen, Nature {\bf 407}, 57 (2000).

\bibitem{Koenig} 
J. K\"onig, J. Schmid, H. Schoeller, and G. Sch\"on, Phys. Rev.
 B {\bf 54}, 16820 (1996).

\bibitem{Meir}
Y. Meir and N.S. Wingreen, Phys. Rev. Lett. {\bf 68}, 2512 (1992).

\bibitem{Hershfield} 
S. Hershfield, J.H. Davies, and J.W. Wilkins, 
Phys. Rev. B {\bf 46}, 7046 (1992).


\bibitem{Abanin05}
D.A.\ Abanin and L.S.\ Levitov, Phys.\ Rev.\ Lett.\
{\bf 94}, 186803 (2005).

\bibitem{Kaminski00}
A. Kaminski, Yu.V. Nazarov, and L.I.
Glazman, Phys. Rev. B {\bf 62}, 8154 (2000).

\bibitem{Mitra05b}
A.\ Mitra, and A.J. 
Millis, Phys.\ Rev.\ B {\bf 76}, 085342 (2007).

\bibitem{Oguri01}
A. Oguri, Phys. Rev. B {\bf 64}, 153305 (2001).  

\bibitem{Aligia}
A.A. Aligia, Phys. Rev. B {\bf 74}, 155125 (2006).

\bibitem{Konik02}
R.M. Konik, H. Saleur, and A. Ludwig, 
Phys. Rev. B {\bf 66}, 125304 (2002).

\bibitem{Rosch04}
A. Rosch, J. Paaske, J. Kroha, and P. W\"olfle, 
Phys. Rev. Lett. {\bf 90}, 076804 (2003).

\bibitem{Fujii}
T.\ Fujii and K.\ Ueda, Phys. Rev. B {\bf 68}, 155310 (2003); 
 J. Phys. Soc. Jpn.  {\bf 74}, 127 (2005).

\bibitem{rubio}
K.S. Thygesen and A. Rubio, Phys. Rev. B {\bf 77}, 115333 (2008).

\bibitem{Komnik} A. Komnik and A.O. Gogolin, Phys. Rev. Lett. {\bf 90},
246403 (2003). 

\bibitem{Doyon1} B. Doyon and N. Andrei, Phys. Rev. B {\bf 73}, 245326 (2006). 

\bibitem{Mehta} P. Mehta and N. Andrei, Phys. Rev. Lett. {\bf 96}, 216802
 (2006). 
 
\bibitem{Borda} L. Borda, K. Vladar, and A. Zawadowski, Phys. Rev. B 
{\bf 75}, 125107 (2007). 

\bibitem{Doyon2}
B. Doyon, Phys. Rev. Lett. {\bf 99}, 076806 (2007). 

\bibitem{Boulat} E. Boulat and H. Saleur, Phys. Rev. B {\bf 77}, 033409 (2008).

\bibitem{Nishino} A. Nishino and N. Hatano, J. Phys. Soc. Jap. 
{\bf 76}, 063002 (2007). 

\bibitem{Schoeller}H. Schoeller and J. K\"onig, Phys. Rev. Lett. 
 {\bf 84}, 3686 (2000).
 
\bibitem{Schmitteckert}
D. Bohr and P. Schmitteckert, Phys. Rev. B {\bf 75}, 241103(R) (2007);
P. Schmitteckert and F. Evers, Phys. Rev. Lett. {\bf 100}, 086401 (2008).
 
\bibitem{Schollwoeck}A.J. Daley, C. Kollath, U. Schollw\"ock, and G. Vidal, 
J. Stat. Mech.: Theor. Exp.  P04005 (2004).
 
\bibitem{Jakobs} S.G. Jakobs, V. Meden, and H. Schoeller, Phys. Rev.
 Lett. {\bf 99}, 150603 (2007). 
 
\bibitem{Costi1} T.A. Costi, Phys. Rev. B {\bf 55}, 3003 (1997). 

\bibitem{anders}
Very recent progress on  the application of the numerical renormalization group method out of equilibrium has been reported by F.B. Anders and A. Schiller, Phys. Rev. B {\bf 74}, 245113 (2006); F.B. Anders, preprint arXiv:0802.0371.

 
\bibitem{Oguri95}A. Oguri, H. Ishii, and T. Saso, Phys. Rev. B {\bf 51}, 4715 
(1995). 

\bibitem{Wang}X. Wang, C.D. Spataru, M.S. Hybertsen, and A.J. Millis, 
Phys. Rev. B {\bf 77}, 045119 (2008).  

\bibitem{Egger00}
C.H. Mak and R. Egger, J. Chem. Phys. {\bf 110}, 12 (1999);
R.\ Egger, L.\ M\"uhlbacher, and C.H.\ Mak, Phys. Rev. E 
{\bf 61}, 5961 (2000).

\bibitem{Egger03}
L.\ M\"uhlbacher and R.\ Egger, J.\ Chem.\ Phys.\ {\bf 118}, 179 (2003). 

\bibitem{lothar}
L. M\"uhlbacher and E. Rabani, Phys. Rev. Lett. {\bf 100}, 176403
(2008).

\bibitem{kehrein}  S. Kehrein, Phys. Rev. Lett. {\bf 95}, 056602 (2005).

\bibitem{Han}
J.E. Han and R.J. Heary, Phys. Rev. Lett. {\bf 99}, 236808 (2007).

\bibitem{Anderson}
P.W. Anderson, Phys. Rev. {\bf 124}, 41 (1961). 

\bibitem{Flensberg} H.\ Bruus and K.\ Flensberg, {\it Many-Body
Quantum Theory in Condensed Matter Physics\/}, (Oxford UP, Oxford, 
2004). 

\bibitem{Tsvelik} A. M. Tsvelik and P. B. Wiegmann, Adv. Phys. {\bf 32}, 453 
(1983). 

\bibitem{Schiller} A. Schiller and S. Hershfield, Phys. Rev. B 
{\bf 51}, 12896 (1995). 

\bibitem{Horvath}
B. Horv{\'a}th, B. Lazarovits, O. Sauret, and G. Zar{\'a}nd,
Phys. Rev. B {\bf 77}, 113108 (2008).

\bibitem{Kamenev} A. Kamenev, in {\it Nanophysics: Coherence and 
Transport\/}, Les Houches session LXXXI, ed.\ H. Bouchiat, 
  Y. Gefen, S. Gu\'eron, G. Montambaux, and J. Dalibard 
(Elsevier, New York, 2005). 

\bibitem{Keldysh} L.V. Keldysh,  Zh. Eksp. Teor. Fiz., {\bf 47}, 
1515 (1964) [Sov. Phys. JETP {\bf 47}, 804 (1961)]. 

\bibitem{Rammer} J. Rammer and H. Smith, Rev. Mod. Phys. {\bf 58}, 323 (1986). 


\bibitem{negele} J.W.\ Negele and H.\ Orland, {\em Quantum
Many-Particle Systems\/} (Addison-Wesley,  Redwood City, 1988). 

\bibitem{HirschFye} J.E. Hirsch, and R.M. Fye, Phys. Rev. Lett. {\bf 56}, 
2521 (1986). 

\bibitem{Hirsch} J.E. Hirsch, Phys. Rev. B {\bf 28}, 4059 (1983).

\bibitem{Fye} R.M. Fye, Phys. Rev. B {\bf 33}, 6271 (1986).

\bibitem{deRaedt}H. De Raedt, and B. De Raedt, Phys. Rev. A {\bf 28}, 3575 (1983).

\bibitem{Eckel}
 J.\ Eckel, S.\ Weiss, and M.\ Thorwart, 
 Eur.\ Phys.\ J.\ B {\bf 53}, 91 (2006).


\bibitem{Hubbard} J. Hubbard, Phys. Rev. Lett. {\bf 3}, 77 (1959).

\bibitem{Siano} F. Siano and R. Egger, Phys. Rev. Lett. {\bf 93}, 047002 (2004).

\bibitem{Weiss} U. Weiss, {\it Quantum Dissipative Systems\/}, 2nd ed.\  
(World Scientific, Singapore, 1999).

\bibitem{Makri}D.E. Makarov and N. Makri, Chem.\ Phys.\ Lett.\
{\bf 221}, 482 (1994); N. Makri and D.E. Makarov, J.\ Chem.\
Phys.\ {\bf 102}, 4600 (1995); N. Makri and D.E. Makarov, J.\ Chem.\
Phys.\ {\bf 102}, 4611 (1995); N. Makri, J.\ Math.\
Phys.\ {\bf 36}, 2430 (1995). 

\bibitem{Rosch05}
A. Rosch, J. Paaske, J. Kroha, and P. W\"olfle, 
J. Phys. Soc. Jap. {\bf 74}, 118 (2005).

\bibitem{Luca}
L. Dell'Anna, A. Zazunov, and R. Egger, Phys. Rev. B {\bf 77}, 104525 (2008).

\bibitem{Costi2}
T.A. Costi, A.C. Hewson, and V. Zlatic, J.\ Phys.: Condens. Matter 
{\bf 6}, 2519 (1994). 

\bibitem{Hamann}
D.R. Hamann, Phys. Rev. {\bf 158}, 570 (1967). 
 
\bibitem{Thorwart1}M. Thorwart and P. Jung, Phys. Rev. Lett. {\bf 78}, 2503
(1997); M. Thorwart, P. Reimann, P. Jung, and R.F. Fox,
Chem.\ Phys.\ {\bf 235}, 61 (1998); M. Thorwart, P. Reimann, P. Jung
and R.F. Fox, Phys. Lett. A {\bf 239}, 233 (1998); 
M. Thorwart, P. Reimann, and P. H\"anggi, 
Phys.\ Rev.\ E {\bf 62}, 5808 (2000); M. Thorwart, E. Paladino, and M.
Grifoni, Chem.\ Phys.\ {\bf 296}, 333 (2004); M.C. Goorden, M. Thorwart, 
and M. Grifoni, Phys. Rev. Lett. {\bf 93}, 267005 (2004); 
M. Thorwart, J. Eckel and E.R. Mucciolo, 
 Phys. Rev. B {\bf 72}, 235320 (2005). 

\bibitem{Feynman} R. Feynman and F.L. Vernon, 
Ann. Phys. (N.Y.) {\bf 24}, 118 (1963).
 

 
\end{thebibliography}
\end{document}